\documentclass[
 reprint,
 amsmath,amssymb,
 aps, superscriptaddress, pre
]{revtex4-1}
\usepackage{graphicx}
\usepackage{amsmath}
\usepackage{amssymb}
\usepackage{enumerate}
\usepackage[justification=raggedright,singlelinecheck=false,font=small]{caption}
\usepackage[justification=raggedright,singlelinecheck=false,font=small]{subcaption}

\def \be{\begin{equation}}
\def \ee{\end{equation}}
\def \ba{\begin{array}}
\def \ea{\end{array}}
\def \bea{\begin{eqnarray}}
\def \eea{\end{eqnarray}}

\begin{document}

%\preprint{APS/123-QED}

\title{Finding stability domains and escape rates in kicked Hamiltonians}% Force line breaks with \\

\author{Archishman Raju}

\author{Sayan Choudhury}
 \affiliation{Laboratory of Atomic and Solid State Physics, Cornell University, Ithaca, New York 14853}%Lines break automatically or can be forced 
\author{David L. Rubin}
 \affiliation{Laboratory of Elementary Particle Physics, Cornell University, Ithaca, New York 14853}
 \author{Amie Wilkinson}
\affiliation{Department of Mathematics, The University of Chicago, Chicago, Illinois 60637}
 \author{James P. Sethna}
  \email{sethna@lassp.cornell.edu}

 \affiliation{Laboratory of Atomic and Solid State Physics, Cornell University, Ithaca, New York 14853}

\date{\today}% It is always \today, today,

\begin{abstract}
We use an effective Hamiltonian to characterize particle dynamics and
find escape rates in a periodically kicked Hamiltonian. We study a model
of particles in storage rings that is described by a chaotic symplectic
map. Ignoring the resonances, the dynamics typically has a finite region
in phase space where it is stable. Inherent noise in the system leads to
particle loss from this stable region. The competition of this noise with radiation damping, which increases stability, determines the escape rate. Determining this `aperture' and
finding escape rates is therefore an important physical problem. We
compare the results of two different perturbation theories and a variational
method to estimate this stable region. Including noise, we derive analytical
estimates for the steady-state populations (and the resulting beam emittance), for the escape rate in the small damping regime,
and compare them with numerical simulations.
%\begin{description}
%\end{description}
\end{abstract}

\maketitle
\section{Introduction}

The study of the physics of nearly integrable systems has a rich and
fascinating history. It has applications in fields varying from
planetary science to accelerator physics. In both accelerators and planetary
motion, the survival of particles under billions of revolutions under a nonlinear
Hamiltonian is subtle; indeed, Hamiltonian chaos was first
discovered~\cite{poincare} in the context of the three-body problem in 
planetary systems. These chaotic resonances have been thoroughly
studied~\cite{arnold,berry,rand}, and cause `small denominator'
problems~\cite{arnol1963small} that prevent otherwise useful perturbative
calculational techniques from converging. Here we shall investigate how
ignoring the chaos -- developing effective integrals of the motion --
can be used to capture the behavior important to the design and optimization
of particle accelerators, and more generally for time-periodic
Hamiltonian systems with islands of long-term stability in phase space.

\begin{figure}[h!]
\begin{center}
    \includegraphics[width=.48\textwidth]{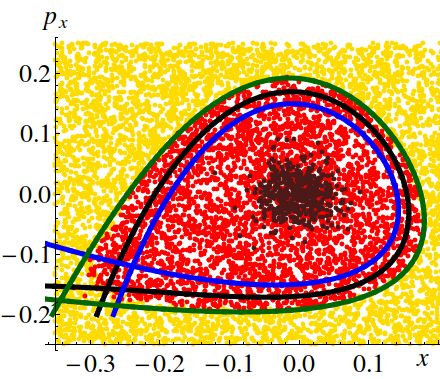}
\end{center}
	\caption{{\bf Phase space regions} for a 1d map (one position,
 one momentum) describing orbits passing through a cross section in an
 accelerator. Yellow points escape to infinity (the walls of the chamber);
 red points are stable for infinite time. We use three methods,
 the normal form method (NF, blue), the Baker-Campbell-Hausdorff expansion
 (BCH, black), and a variational method (VAR, green)
 to estimate the {\em aperture} of stable orbits.
 The black points represent the {\em bunch} in phase space formed at
 long times when particles are subject to noise and damping. The
 phase-space extent of this bunch is the {\em emittance} which
 characterizes the brightness of the beam. Our methods can also estimate
 the rate of escape of bunch particles from the aperture (not shown).}
     \label{fig:summaryAperture}
\end{figure}

Fig.~\ref{fig:summaryAperture} provides an illustration of our results.
In section~\ref{sec:MapsInvariants} we introduce traditional toy models for
the dynamics of accelerators, and three strategies for calculating
approximate invariants: the normal form method (NF),
the Baker Campbell Hausdorff (BCH) expansion, and a variational method. In
section~\ref{sec:Aperture} we use the three invariants (curves in
Fig.~\ref{fig:summaryAperture}) to approximate the
`aperture' -- the region in phase space where lifetimes are effectively
infinite. In section~\ref{sec:Emittance} we use our invariants to describe the
equilibrium distribution of particles in a noisy environment, allowing
the characterization of emittance (the phase-space volume of the bunch). 
In section~\ref{sec:Escape} we generalize theories of chemical reaction
rates to estimate the particle loss rate in the accelerator, and also 
provide a controlled, corrected form for the equilibrium distribution.
These calculations, while useful in this context, are also generally applicable for kicked noisy maps. 
Our analysis in sections~\ref{sec:Emittance} and~\ref{sec:Escape} is
confined to a 1d map, but higher dimensional systems are discussed both
to motivate our approximations and to outline how our methods could be 
generalized.

Our methods will bypass the chaotic resonances that have fascinated
mathematicians and dynamical systems theorists in the last century. 
The aperture of stable orbits for maps of more than one dimension
is mathematically a strange set, presumably with an open dense set of holes
corresponding to chaotic resonances connected by Arnol'd
diffusion~\cite{arnold1964instability}. The fact that accelerator designers
characterize their apertures as simple sets motivates our use of
invariants. Our approximate invariants ignore these holes, except 
insofar as resonances determine the outer boundary of stability. 
Designers avoid strong resonances, facilitating the use of 
our methods. Our focus, therefore, is on accurate estimates of the qualitative
stability boundary, and on calculating the resulting emittance and escape
rates in the presence of noise.

%However, the existence of islands of stable orbits
%separated by chaos seems to be common, and deserves a physical or
%mathematical description.

\section{Map and Invariants}
\label{sec:MapsInvariants}

Particle orbits in accelerators are often well represented by a Poincar{\'e} recurrence maps. These maps usually describe nearly harmonic systems with nonlinearities coming from sextupole and other higher order magnets. Most trajectories near the central `reference orbit' are stable for infinite time ({\em i.e.}, live on KAM tori~\cite{kolmogorov1954conservation,moser1962invariant,arnold1963proof}); at farther distances where the nonlinearities are large orbits escape to hit the
chamber walls. In accelerators it is found that the region of practically stable orbits is well described as a simple region called the `dynamic aperture' (sometimes surrounded by a cycle of islands with similar properties). We shall refer to the stable region in phase-space as the  `aperture' in this paper. 

%The maps give the phase space coordinates of a particle after a fixed time step. We show the cross-sections of one such map in 1d in Figure~\ref{fig:summaryAperture}. The figure is plotted by taking initial conditions, evolving them with the map for a fixed number of turns, and coloring points green if they escape to infinity or red otherwise. The figure makes it clear that there is a stable region in phase space separated from the unstable region by an aperture. 

In practice, this aperture is often found numerically by running the map for different initial conditions and seeing which particles escape. Here, we use two different kinds of perturbation theory, the normal form method (NF) and the Baker Campbell Hausdorff (BCH) expansion to try and estimate the aperture. We also use a variational method that improves on both of these methods. Our general strategy is to find one or more approximate invariants of the map and find its saddle points. The contour at one of the saddle points gives our approximation to the aperture. 

\begin{figure}[h!]
\begin{center}
    \includegraphics[width=.35\textwidth]{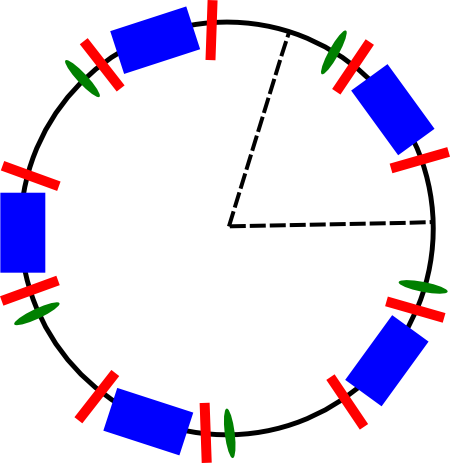}
\end{center}
	\caption{The figure above shows the toy accelerator ring that we model in this paper. The ring has \textit{linear elements} which are the dipole (in blue) and quadrupole (in red). These can be represented as a harmonic oscillator. The nonlinear sextupole (in green) provides a kick at periodic intervals. Our calculations here work for a periodic array, and our map models the section between the dashed lines for a particle moving counter-clockwise. In any real accelerator, the sextupoles and other elements would have different strengths along the ring. Our methods would still work but the actual calculations would be messier.}
     \label{fig:ring}
\end{figure}

As our toy example in 1-d, we will study a harmonic Hamiltonian with a kick, 
\begin{equation}
\label{Ham}
 H = \frac{p^2}{2 m} + \frac{m \omega^2 x^2}{2} + \frac{K x^3}{6} \sum_n \delta(t - n \tau) .
\end{equation}
Here, $\omega$ is the frequency of the particle (as it wiggles perpendicular to the direction of motion) and $\tau$ is the period between kicks. Here, the form of the kick models the action of sextupole magnets in particle accelerators~\cite{bazzani1988normal}. Henceforth, we will set $m = 1$. The dynamics then corresponds to the classic H\'{e}non map 
\begin{align}
 x_{n+1}  &= x_n \cos \omega \tau + \frac{p_n}{\omega} \sin \omega \tau , \\
 p_{n+1} &= p_n \cos \omega \tau - \omega x_n \sin \omega \tau - \frac{K}{2} x_{n+1}^2 .
\end{align}
 Such kicked systems have become a paradigmatic example of chaos in both classical and quantum systems. In accelerators, one often only has access to the map and not to the original time-dependent Hamiltonian. We will formally denote the linear part of map without the kick by $\mathcal{M}$. We will denote the action of the kick, the non-linear part by $\mathcal{K}$. So, $\mathcal{M}(x_n) = x_{n+1}, \mathcal{K}(\mathcal{M}(p_n)) = p_{n+1}$. When acting on some function $f$ of $x_n$ and $p_n$, we have $\mathcal{M}(f(x_n, p_n)) = f(\mathcal{M}(x_n), \mathcal{M}(p_n))$.

%We also define the dimensionless parameter $b = \omega \tau$ and the length $l = \frac{m \omega^2 \tau}{K}$. Classical perturbation theory is in $K$ or equivalently in $\frac{1}{l}$.

We note that the aperture is not a simple region in any dimension greater than one. Yet, we are inspired to use perturbation theories which give approximate invariants to capture simple regions which remain stable in practice. There is a large literature on constructing invariants for nonlinear systems. For symplectic maps, the NF~\cite{birkhoff1960dynamical, moser1968lectures} is the most commonly used method. The NF gives approximate invariants up to a certain order in $K$ but fails to capture the resonances. Resonant NF theory~\cite{bazzani1993resonant}
can be used to include the effect of the resonances. Lie-algebraic techniques, which include the BCH expansion, can also be used to calculate invariants~\cite{lie1, lie2, lie3, chaos}. Finally, people have tried numerical non-perturbative variational methods to capture the aperture by fitting polynomials or Fourier coefficients of generating functions to trajectories \cite{approx1, approx2, approx3}. Many of these methods usually concentrate on getting the detailed structure of the aperture, including the islands, often at the cost of getting the boundary correctly~\cite{servizi1992resonant}. 

Since accelerators are designed to avoid these large resonances, our focus is on getting accurate estimates of the stability boundary. As we will see later, this is particularly crucial to calculate the escape rate. It is known that both the NF and BCH lead to asymptotic series~\cite{bazzani1989nekhoroshev, ScharfBCH}. Traditionally, the way to make sense of higher order terms in asymptotic series is by resumming them~\cite{hardy2000divergent}. It is conceivable that detailed studies of a single resonance and the interaction between multiple resonances~\cite{chirikov1979universal} could be used to create a resummation method which both captures the effect of the resonances and gives an accurate stability boundary.

%[Numerical variational method here.]

%[Explain physically why can't have chaos and complete set of NF invariants.
%Explain physically why can't have effective Hamiltonian w/chaos. Note
%complication: there is a spatially dependent time-independent %Hamiltonian
%if we ignore the AC accelerating cavities. This leads to 5D map vs 6D map?
%Refer to many math papers about this. Asymptotic expansion theorems. Variational
%methods.]

 While there are an infinite number of functions which can serve as approximate invariants (because any function of an invariant is invariant), one natural approach is to construct an effective Hamiltonian which also captures the dynamics of the system. Periodically driven systems are a class of time-dependent Hamiltonian systems for which an effective time-independent Hamiltonian (and consequently the aperture) can be obtained by an exact analytical formalism known as the Floquet formalism. The Floquet formalism is well known in classical accelerator physics~\cite{edwards2008introduction} but the concept of a Floquet Hamiltonian is best understood using quantum mechanics. Thus in the spirit of Ref~\cite{polkovnikov2014floquet}, we first obtain the effective Hamiltonian for the quantum version of the Hamiltonian given  by Eq.(\ref{Ham}) and then take the classical limit.  The Floquet formalism involves calculating the evolution operator after n periods $U(nt)$ which is given by :
\begin{equation}
U(nT) = \mathcal{T}  \exp \left(-i \int_0^{nT} H(t)/\hbar\right) = U(T)^n .
\end{equation}
Thus, the evolution operator for 1 period is defined by :
\be
U(T) = \mathcal{T}  \exp \left(-i \int_0^{T} H(t)\right) = \exp(-i H_{\rm eff} T/\hbar) ,
\ee
where $H_{\rm eff}$ is the effective Hamiltonian and $\mathcal{T}$ is a time-ordering operator. For the Hamiltonian in Eq.(\ref{Ham}), this equation is particularly simple and we obtain :
\be
U(T) = \exp \left(-i \frac{K x^3}{6 \hbar}\right) \exp\left(-i(\frac{p^2}{2} + \frac{\omega^2 x^2}{2})\tau \right) .
\ee
The effective Hamiltonian we get using Floquet theory, we will call $H_{\rm BCH}$. It is given by :
\be
H_{\rm BCH} = i \hbar \log \left(\exp \left(-i \frac{K x^3}{6 \hbar}\right) \exp\left(-i(\frac{p^2}{2} + \frac{\omega^2 x^2}{2})\tau \right)\right) .
\ee
Now employing the Baker-Campbell-Hausdorff expansion and going to the classical problem in the usual way~\cite{ScharfBCH}, we obtain the effective Hamiltonian (up to second order):
\begin{widetext}
\be
\label{effectiveham}
H_{\rm BCH} = \frac{1}{2} \left(x^2 \omega ^2+p^2 \right) + \frac{K^2 \tau  x^4-2 K x \left(x^2 \left(\tau ^2 \omega ^2-4\right)-6 p \tau  x-2 p^2 \tau ^2\right) }{48 \tau } + O(\tau^3) .
\ee
\end{widetext}
One major difference between the classical and quantum problems is that an effective Hamiltonian always exists for the quantum case, but the effective Hamiltonian description breaks down near resonances for the classical case. As has been argued in~\cite{ScharfBCH}, for the quantum problem, the Baker-Campbell-Hausdorff expansion also breaks down near resonances, even though an effective Hamiltonian exists.

The more commonly used perturbation theory is the normal form method. In the context of canonical systems, this is called the Birkhoff Normal Form. Here, we do not construct canonical transformations which take the Hamiltonian to a normal form but instead directly construct (multiple) invariants of the map. The essence of the normal form method is to convert the problem of finding an invariant to a linear algebra problem on the space of homogeneous polynomials. This can be done by noticing that the action of $\mathcal{K}$ on any polynomial is to increase its order by 1. Let us start with the invariant of the linear part of the map $\mathcal{M}$ which is just the second-order polynomial $I_2 = \omega^2 x^2 + p^2$. Now, if we choose a third order polynomial $I_3$ so that the action of $\mathcal{M}$ on $I_3$ exactly cancels the action of $\mathcal{K}$ on $I_2$, we get an approximate invariant up to third order. We can continue this process order-by-order to get higher and higher order approximate invariants. The NF Hamiltonian to 3rd order is given by 
\begin{widetext}
\begin{equation}
 H_{\rm NF} = \frac{1}{2} \left(p^2+x^2 \omega ^2 + \frac{K p^2 x \sin (\omega \tau )}{2 \omega +4 \omega  \cos (\omega \tau
   )}+\frac{1}{2} K p x^2+\frac{K x^3 \omega \tau 
   \left(\left(\cos ^3(\omega \tau )+1\right) \cot (\omega \tau)+\sin (\omega \tau )
   \cos ^2(\omega \tau )\right)}{4 \cos (\omega \tau)+2}\right) + \mathcal{O}(K^2).
\end{equation}
\end{widetext}
%{\bf [XXX Say if this the order we use for the plots. Mention here that 
%fifth order (if I remember right) works best, and sixth order loses the 
%saddlepoint. Offer supplemental material or a Web site or something giving
%the BCH and NF invariants to other orders.]}

The fourth order NF Hamiltonian loses the saddle point. All expansions have been truncated at the order which best describes the aperture (see supplementary material for details). The NF and BCH Hamiltonians are both time-independent Hamiltonians trying to capture the one-period dynamics of the map. One method perturbs in the nonlinearity and the other perturbs in the inverse-frequency of the kick. If the series generated by perturbation theory were to converge, both would converge to the same effective Hamiltonian. However, because the series are asymptotic, they are most useful when truncated to a low order. The effectiveness of such low order truncations will depend on the particular problem. Finally, we can improve on the estimates of both of these perturbative methods numerically. One way to do this is to start from a point that lies on the aperture predicted by perturbation theory, and generate a trajectory using the map. Inspired by the form of the Hamiltonians obtained using perturbation theory, one can then simply fit a fourth order polynomial whose quadratic terms are constrained to be $p^2/2 + x^2 \omega^2/2$ to the trajectory. The fit is generated by minimizing the variation of this polynomial over 10000 iterates of the map. The variational Hamiltonian we obtain for parameter values $\tau = 1$, $K = 6$, $\omega = 0.96$ is given by
\begin{widetext}
\begin{equation}
 H_{\rm VAR} = \frac{p^2}{2}+\frac{1}{2} x^2 \omega ^2 + a_1 x^3+ a_3 p x^2+ a_4 p^2 x+ b_1 x^4+ b_3
   p x^3+ b_5 p^2 x^2 ,
\end{equation}
\end{widetext}
with best fit parameters $a_1 = 0.73, a_3 = 1.47, a_4 = 0.56, b_1 = 0.34, b_3 = 1.15, b_5 = 0.45$.

\section {Aperture}
\label{sec:Aperture}

\begin{figure}[ht]
\centering
  \begin{subfigure}[b]{.49\linewidth}
    \centering
    \includegraphics[width=.99\textwidth]{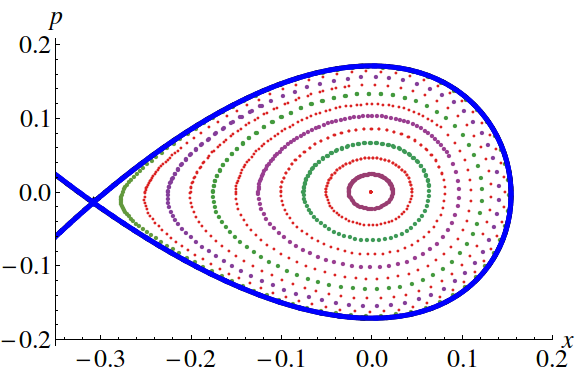}
    \caption{}\label{meanfielda}
  \end{subfigure}%   
  \begin{subfigure}[b]{.49\linewidth}
    \centering
    \includegraphics[width=.99\textwidth]{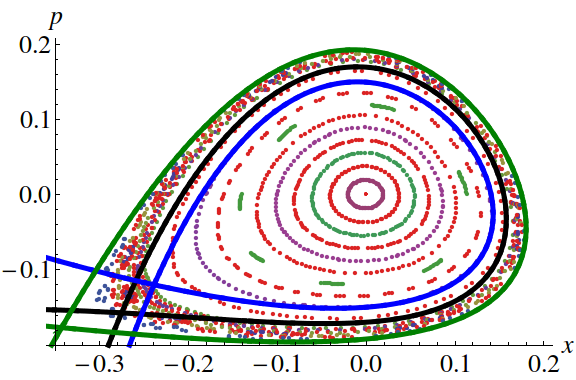}
    \caption{}\label{meanfieldb}
  \end{subfigure}\\% 
	\caption{A plot of the aperture obtained using the NF(blue line), BCH (black line) and numerically (green line) on the Poincar\'{e} cross-section of the dynamics generated by the map. Parameters used here are (a) $\tau = 0.1$, $K = 0.6$, $\omega = 0.96$, (b) $\tau = 1$, $K = 6$, $\omega = 0.96$. In case (a), the two perturbation theories gives results which work well and are practically indistinguishable.} 
\label{dynamicaperture}
\end{figure}

In order to obtain the aperture from the Hamiltonian (or any invariant) we obtain its saddle point. This is given by simultaneously solving the equation ${\partial H_{\rm eff}}/{\partial x} = 0$ and ${\partial H_{\rm eff}}/{\partial p} = 0$. The energy contour corresponding to the saddle point gives the aperture. Examples of the aperture that we obtain for different parameters are shown in Fig~\ref{dynamicaperture}. We show a comparison between the results of the BCH expansion, the NF and our numerical fit.    

\begin{figure}[ht]
\centering
  \begin{subfigure}[b]{.49\linewidth}
    \centering
    \includegraphics[width=.99\textwidth]{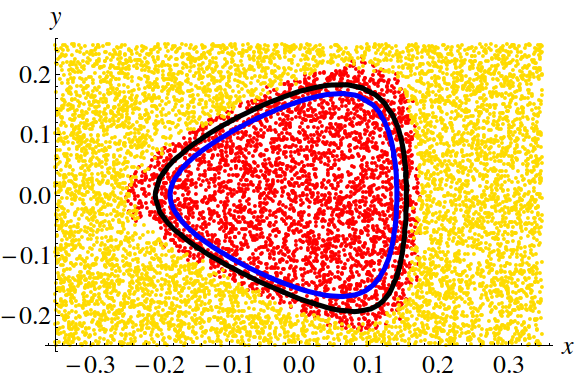}
    \caption{}\label{dyn2da}
  \end{subfigure}%   
  \begin{subfigure}[b]{.49\linewidth}
    \centering
    \includegraphics[width=.99\textwidth]{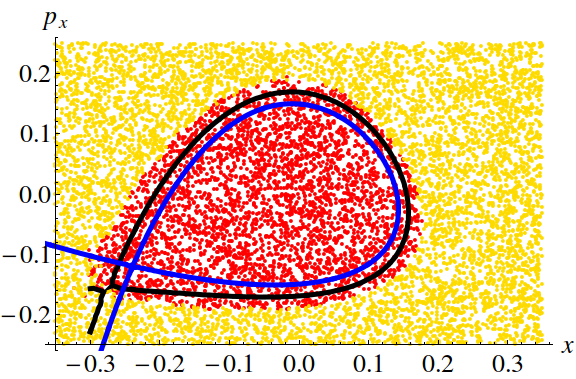}
    \caption{}\label{dyn2db}
  \end{subfigure}\\% 
	\caption{The figure shows two cross-sections of the map in $p_x-y$ plane and the $p_x-x$ plane. The 2-d map used is a generalization of the 1-d map and is given in the supplementary material. The yellow points are initial conditions which escape after a fixed number of turns, while the red points remain bounded. We use two perturbation theories, NF (in blue) and BCH (in black) to estimate the boundary between the two (see text).}
\label{dynamic2d}
\end{figure}

\begin{figure}[ht]
\begin {center}

		\includegraphics[scale = 0.3]{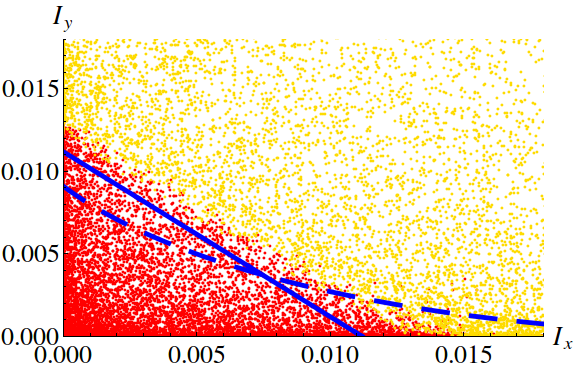}
	\caption{A plot of the two approximate invariants $I_x$ and $I_y$ obtained using the NF. Red points are initial conditions which stay bounded and yellow points are those which escape. The solid blue curve is the energy contour given by $I_x + I_y = c$ where $c$ is the saddle point energy of $I_x+I_y$. The dashed blue curve is plotted by holding one invariant constant and finding the saddle of the other.}
\label{invariantscurve}
\end {center}
\end{figure}

We generalize the map to two dimensions by adding a harmonic oscillator in the $y$ variable and including a kick of the form $K x y^2/2$ in the $y$ momentum~\cite{giovannozzi1998study}. The form of the kick again is a sextupole, and is chosen to satisfy Laplace's equations. We can use the BCH expansion in exactly the same way to construct an effective Hamiltonian. The NF, on the other hand, gives multiple invariants in higher dimensions. The aperture we obtain is shown in Figure~\ref{dynamic2d}. In two dimensions, the NF gives two invariants for this map. The aperture then becomes a curve in the space of the two invariants. In Figure~\ref{invariantscurve}, we show a plot of the initial conditions that escape to infinity (in yellow) and those which stay bounded (in red) in the space of two approximate NF invariants. The curve which sets the boundary can be found by fixing a value of one of the invariants, and finding the constrained saddle point of the other. This problem can be solved using a Lagrange multiplier (see supplementary information) and the solution is shown as a dashed blue line in Figure~\ref{invariantscurve}. It is clear that this curve is not a very good approximation of the boundary.

As comparison, we also show the boundary that we get by simply adding the two invariants to get an approximate `energy' shown using the solid blue line. Remarkably, the approximate `energy' given by this linear combination of the two invariants seems to represent the geometry of the problem better than the blue curve. It is possible that the aperture in higher dimensions is controlled only by the effective Hamiltonian obtained by simply adding the two invariants. Indeed, this is the combination we use to find the NF aperture in Figure~\ref{dynamic2d}. This NF effective Hamiltonian is the analogue of our BCH effective Hamiltonian in two dimensions.

\section{Noise and Emittance}
\label{sec:Emittance}
The effective Hamiltonian can be used to calculate the aperture but it is also useful to study the effect of noise on the dynamics. We are inspired to extend calculations done in the context of chemical reactions to describe particles escaping stability boundaries. There are many sources of noise in accelerators. These include residual gas scattering~\cite{moller1999beam}, photon shot noise~\cite{jowett} and intra-beam scattering~\cite{piwinski1988intra}. Each of these have different forms but they all have the effect of changing the phase-space coordinates of the particle. We will only model the particle loss that occurs as a result of the particles crossing the barrier set by the dynamic aperture (other sources of particle loss exist in real accelerators). We will model the noise phenomenologically with uncorrelated Gaussian noise and linear damping. This assumption has been used earlier to model noise in accelerators~\cite{fokkerplanck1, fokkerplanck2}. A more realistic treatment of the noise would be multiplicative and could be added in principle though some parts of the calculation will then have to be done numerically.  For time-independent Hamiltonian dynamics with a barrier, the effect of noise is well known at least since Kramer, who used the flux-over-population method to calculate the escape rate of particles both in the strongly damped and weakly damped case. There has been some work on the escape rate for maps in the strong damping regime~\cite{reimanntalkner}. The noise, whatever its form, cannot be directly added to the effective Hamiltonian that we calculate in the previous section and must instead be added to the original dynamics. We will do this here only for one dimension (one position and one momentum
in the map). For every time period, before the kick, the equations of motion then are 
\begin{align}
 \dot{x} &= p , \\
 \dot{p} &= -\omega^2 x - \gamma p + \xi(t) . 
\end{align}
We take the noise $\xi(t)$ to be delta-correlated Gaussian noise specified by its two-point function $\xi(t) \xi(t') = 2 \gamma T \delta (t - t')$.  Integrating the equations of motion over one time period and adding the kick gives the noisy map 
\begin{widetext}
\begin{align}
\label{noisymap}
x_{n+1} &= e^{-\frac{\gamma  \tau }{2}} \left(\frac{(2 p_n+\gamma 
   x_n) \sin (\tau  \omega_r )}{2 \omega_r }+x_n \cos (\tau 
   \omega_r )\right) + \xi_{Xn} , \\
   \label{noisymap2}
p_{n+1} &= e^{-\frac{\gamma  \tau }{2}} \left(p_n
  \left(\cos (\tau  \omega_r )-\frac{\gamma  \sin (\tau  \omega_r )}{2
   \omega_r }\right)-\frac{x_n \left(\gamma ^2+4 \omega_r^2\right) \sin (\tau  \omega_r )}{4 \omega_r }\right) - K \frac{x_{n+1}^2}{2} + \xi_{Pn} ,
\end{align}

where the integrated noise terms have zero mean and correlation functions 
\begin{align}
\langle \xi_{Pn} \xi_{Pm} \rangle &= 
\frac{e^{-\gamma  \tau } \left(\gamma ^2 \cos (2 \tau  \omega_r)-\gamma ^2+2 \gamma  \omega_r \sin (2 \tau  \omega_r)-4 \omega_r^2\right)+4 \omega_r^2}{4 
   \omega_r^2} T \delta_{n m}  , \\
   \langle \xi_{Xn} \xi_{Xm}  \rangle &= \\ \nonumber &\frac{4 \omega_r^2-e^{-\gamma  \tau } \left(\gamma ^2 (-\cos (2
   \tau  \omega_r))+\gamma ^2+2 \gamma  \omega_r \sin (2
   \tau  \omega_r)+4 \omega_r^2\right)}{\omega_r^2 \left(\gamma ^3+4 \gamma  \omega_r^2\right)} \gamma T \delta_{n m} , \\
  \langle \xi_{Xn} \xi_{Pn} \rangle &= 
\frac{e^{-\gamma  \tau } \sin ^2(\tau  \omega_r)}{\omega_r^2} \gamma T \delta_{n m}
\end{align}
and $\omega_r^2 = \omega^2 - \frac{\gamma^2}{4}$. 
\end{widetext}

\begin{figure}[ht]
\begin {center}

		\includegraphics[width = 0.45 \textwidth]{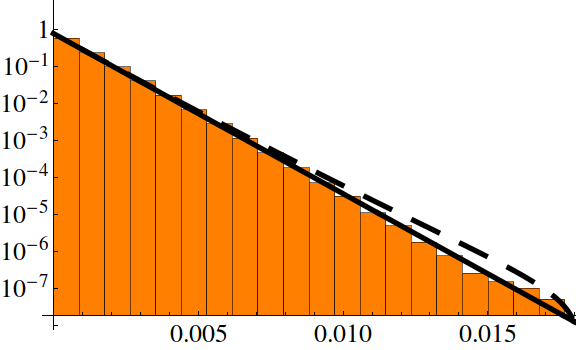}
	\caption{A histogram of the effective energy on a logarithmic scale of the particles which do not escape shows that a Boltzmann distribution in effective energy (solid line) given by our variational Hamiltonian serves as a good approximation to the equilibrium distribution. It is interesting to note that an improved estimate of the equilibrium distribution (dashed line) using Equation~\ref{equilibdist} actually does a worse job of capturing the numerical statistics. This might be because our variational Hamiltonian does not capture the dynamics (and resonances) accurately. Our escape rate calculations do not actually utilize the full form of the distribution because of the approximations we make. Parameter values used here in the simulation are $\tau = 1$, $K = 6$, $\omega = 0.96$, $\gamma = 0.005$, $T = 0.001$. We show the comparison to a typical harmonic approximation of the Hamiltonian in the supplementary material.}
\label{energyhistogram}
\end {center}
\end{figure}

We can use our effective Hamiltonian to calculate the spread of the particle bunch in phase space when noise is added to the map. As we show in Figure~\ref{energyhistogram}, a Boltzmann distribution with a temperature in the effective distribution does a good job of describing the equilibrium distribution of the particles near the center. This is only an approximation to the true equilibrium distribution which we will calculate in the next section. The fact that accelerator designers characterize their bunches with effective temperatures for vertical and horizontal directions~\cite{sands1970physics} is one motivation for the development of invariants that act as vertical and horizontal Hamiltonians, weakly coupled by noise in 2-d.

\section{Steady State and Escape Rate}
\label{sec:Escape}
Equations~(\ref{noisymap}-\ref{noisymap2}) are completely general. To make progress, we make the assumption that we are in the weak damping regime ($\gamma$ small or $1/\gamma$ large compared to all other time-scales in the system). This is usually a realistic assumption for storage rings~\cite{sands1970physics}. Henceforth, we will work only to linear order in $\gamma$. We can then calculate the slow diffusion of the effective Hamiltonian under the noisy dynamics. This diffusion takes place on the Poincar\'{e} section. Hence, we've converted a non-equilibrium problem to an equilibrium problem on the Poincar\'{e} section. 
(The procedure fails in the resonant regions near where the frequencies are rationally related; transport across these resonances is dominated by chaos and
escape rates from islands. Our numerics here are partially testing whether
ignoring these resonances is valid.)
Calculating the escape rate and the steady state probability distribution requires us to first know the drift and diffusion coefficient of the effective Hamiltonian. To find this, we change variables in the usual way \cite{risken}
\begin{align}
\label{variables}
 D_E (x, p) &= \sum_{\alpha = x, p} \frac{\partial H }{\partial \alpha} D_{\alpha} + \sum_{\alpha, \beta = x, p} \frac{\partial^2 H}{\partial \alpha \partial \beta} D_{\alpha \beta} , \\
 D_{E E} (x, p) &= \sum_{\alpha, \beta = x, p} \frac{\partial H}{\partial \alpha} \frac{\partial H}{\partial \beta} D_{\alpha \beta} .
\end{align}
The diffusion coefficients in $x, p$ are defined using the correlation coefficients we calculated earlier. So, for example, $D_{x x} = \frac{\langle \xi_{Xn} \xi_{Xn}  \rangle}{2 \tau}$. There is a slight subtlety in making these change of variables because of the fact that our slow variable is $\gamma$. Using the notation developed earlier, we note that the difference in the Hamiltonian evaluated after one period is given by 
\begin{equation}
\begin{aligned} 
\Delta H =& H\big(\mathcal{M}(x_n) + \xi_{Xn}, \mathcal{K}(\mathcal{M}(p_n) + \xi_{P_n})\big) \\
	&~~ - H\big(x_n, p_n\big) \\
=& H\big(\mathcal{M}(x_n) + \xi_{Xn}, \\
  &~~~~~\mathcal{K}(\mathcal{M}(p_n)) + \xi_{P_n} - K \mathcal{M}(x_n) \xi_{Xn} \big)  \\
 &- H\big(x_n, p_n\big)  \\
=& H\big(\mathcal{M}(x_n), \mathcal{K}(\mathcal{M}(p_n))\big) - H\big(x_n, p_n\big) \\
& + \frac{\partial H}{\partial x} \xi_{X_n}  
	+ \frac{\partial H}{\partial p} (\xi_{Pn} - K \xi_{Xn} \mathcal{M}(x_n))   \\
=& \frac{\partial H}{\partial x} \xi_{Xn} + \frac{\partial H}{\partial p} (\xi_{Pn} - K \xi_{Xn} \mathcal{M}(x_n)),
\end{aligned}
\end{equation}
where we have kept only linear terms in $\gamma$ (and not written the second order contribution to the drift). We have also ignored the higher order terms of the perturbation theory in the effective Hamiltonian and assumed it to be a faithful characterization of the dynamics of the map. Note that the partial derivatives must be evaluated at the new points of the unkicked map and the fact that the kick takes place after the evolution requires us to be more careful about the noise in the $p$ direction.  

\begin{figure*}[ht]
\centering
  \begin{subfigure}[b]{.49\linewidth}
    \centering
    \includegraphics[width=.99\textwidth]{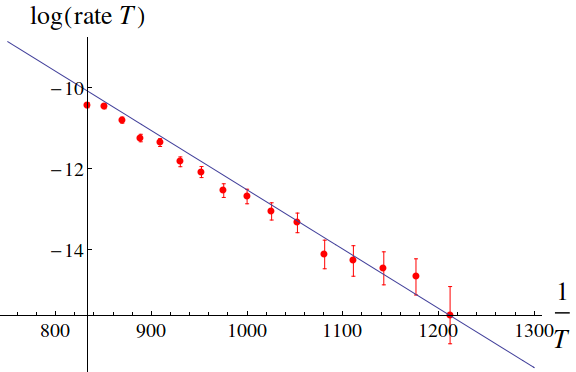}
    \caption{}\label{tau01}
  \end{subfigure}%   
  \begin{subfigure}[b]{.49\linewidth}
    \centering
    \includegraphics[width=.99\textwidth]{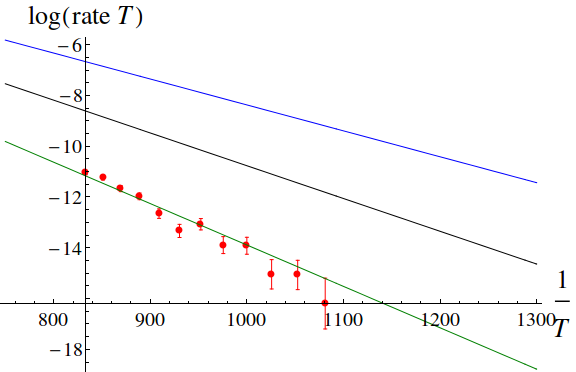}
    \caption{}\label{tau1}
  \end{subfigure}\\% 
	\caption{We compare our analytical predictions for the escape rate with numerical results. Because of the form of the rate given in Equation~\ref{rateeq}, we can plot $\log (k T)$ vs $1/T$ to get a straight line. Error bars are drawn from Poisson statistics. (a) In the case where perturbation theory works, with parameter values $\tau = 0.1, K  = 0.6, \gamma = 0.005$, the effective Hamiltonian obtained from either perturbation theory does a good job of capturing the aperture. (b) Here, parameter values are $\tau = 1, K = 6, \gamma = 0.005$. Whereas both the effective Hamiltonian generated from BCH (black line) and from NF (blue line) fail to capture the escape rate (mostly because they get the wrong $E_b$), we show that one can use a variational method to improve the estimate of the barrier energy and get a good estimate of the escape rate (green line)  }
\label{straightlinetau1}
\end{figure*}

These drift and diffusion coefficients have to be averaged over the other canonical variable which acts as a time coordinate for the effective Hamiltonian. Calling this variable $s$, we then see that. 
\begin{align}
 D_E (E) &= \frac{1}{S} \oint D_E(x, p) ds , \\
 &= \frac{1}{S} \oint\frac{D_E(x, p)}{\frac{\partial H}{\partial p}} dx , \\
 D_{E E} (E) &= \frac{1}{S} \oint\frac{D_{E E}(x, p)}{\frac{\partial H}{\partial p}} dx ,
\end{align}
where $S = \oint ds $. Even though an exact analytical expression for $D_{E E}(x, p)$ is easy to calculate using standard computer-algebra software, these integrals have to be typically evaluated numerically. Having averaged over the fast variable, we can now make a stochastic differential equation using the prescription
\begin{equation}
 \frac{d E}{dt} = D_E(E) + \xi_{E} (t) , 
\end{equation}
where $\langle \xi_{E} (t) \xi_{E} (t') \rangle = 2 D_{E E} (E) \delta (t - t')$. We now go from a Langevin equation to a Fokker-Planck equation in the usual way.
\begin{equation}
 \frac{\partial \rho}{\partial t} = -\frac{\partial}{\partial E} (\rho D_E) + \frac{\partial^2}{\partial E^2} \rho D_{E E}.
\end{equation}
The solution to this equation with a steady state current, a source at $E = 0$ and a sink at the barrier energy gives the approximate equilibrium probability distribution. This distribution is a Boltzmann distribution to linear order in the energy with higher order corrections in $E$.
%The approximate equilibrium probability distribution is characterized by a state of zero flux and can be obtained through this equation
%\begin{equation}
% -\rho D_E + \frac{\partial \rho}{\partial E} D_{EE} + D^\prime_{EE} \rho = 0
%\end{equation}
%or
%\begin{equation}
% \rho = e^{\int \frac{-D_E + D^\prime_{EE}}{D_{EE}} dE}.
%\end{equation}
%The integral in this expression goes linearly with $E$ to lowest order. Therefore, this distribution is a Boltzmann distribution with higher order corrections in $E$. 

We now use the flux-over population method to solve for the escape rate. The flux-over population method involves solving the above equations with a constant steady state current and dividing by the density to find the escape rate~\cite{reactionratetheory}. That is, we want to solve the differential equation
\begin{align}
 -D_E (E) \rho + \frac{\partial}{\partial E} (D_{E E} \rho ) &= J , \\
 \frac{-D_E + D^\prime_{EE}}{D_{EE}} \rho + \frac{\partial \rho}{\partial E} &= \frac{J}{D_{EE}} , 
\end{align}
Define 
\begin{align}
 v(E) &= \int \frac{-D_E + D^\prime_{EE}}{D_{EE}} dE , \\
 &= \int -\frac{D_E}{D_{EE}} dE + \log D_{EE} .
\end{align}
Then it can be checked that the solution is 
\begin{equation}
\label{equilibdist}
 \rho(E) = e^{-v(E)} \int \frac{e^{v(E)}}{D_{EE}} J dE .
\end{equation}
This equation gives the full form of the equilibrium probability distribution. The inverse of the escape rate is given by $k^{-1} = \frac{\int_0^{E_b} \rho(E)}{J}.$ This simplifies to  
\begin{equation}
 k^{-1} = \int_0^{E_b} \frac{e^{\int \frac{D_E}{D_{EE}}}}{D_{EE}} dE \int_E^{E_b} e^{\int -\frac{D_E}{D_{EE}}} dE' .
\end{equation}
The first integral (over $E'$) is dominated by its value at $E_b$. The second integral (over $E$) is dominated by its value at $0$. Estimating the integral by its value at these boundaries gives us a first approximation to $k$ (which is valid for $E_b \gg T$). Doing this requires us to evaluate the integral $\int D_E/D_{EE} dE $. Going back to Equation~\ref{variables}, we see that the drift coefficient has two terms. The first term is independent of $T$ while the second term is linear in $T$. Hence, the above integral has a part which depends on $T$ and contributes to the exponent. This is well behaved everywhere. The other part contributes to the pre-factor and has a singularity at $E = 0$ because the diffusion coefficient vanishes there. Hence estimating the escape rate requires one to find the finite limit $e^{-\int D_E/D_{EE}}/D_{EE}$ converges to at $0$. This was done numerically by evaluating the integral for finite $\epsilon$ and then taking $\epsilon$ very small. In general, this means the escape rate has the form 
\begin{equation}
\label{rateeq}
 k = \frac{a_0 \gamma}{T} e^{-f(E_b)/T} ,
\end{equation}
where $a_0$ is a pre-factor and $f(E_b)$ is some function of the barrier. Both of these are independent of the damping and the temperature.

We note that the escape rate calculation is independent of the perturbation theory used to generate the Hamiltonian. Since it depends exponentially on the energy barrier, it is very sensitive to the position of the barrier. We show a comparison of the prediction of the escape rate with simulations in Figure~\ref{straightlinetau1}. 

Calculating the escape rate in higher dimensions is more complicated. In 2-d, there are two slow variables given by the invariant in the vertical and horizontal direction. The noise in different directions is typically very different~\cite{sands1970physics}. There are three possible approaches which we will explore in future work. One is to derive and solve the full diffusion equation in 2-d. Secondly, one can derive the decay rate in the limit where the coupling between the two directions is small. Finally, we can solve the full non-equilibrium problem using methods like transition path theory \cite{maier1993escape} which were developed to deal with inherently non-equilibrium systems.

\section{Conclusion}

%XXX Explain more mathematical description better
We have here compared two different perturbation theories and numerically improved them to calculate the aperture of a harmonic map with a nonlinear kick. We have also calculated the emittance and escape rate in 1-d. Our method is expected to work for systems without dangerous resonances and weak damping.  A lot of effort has gone into understanding the resonances which inevitably prohibit the presence of a simple aperture.  However, the existence of relatively stable `islands' of KAM tori in phase space, embedded in a sea of unstable, chaotic trajectories, is a commonly observed phenomenon. Here our aperture is such an island, and our exploration of techniques for calculating it is a special case of a general mathematical challenge.

%While we have explored calculational techniques to characterize the stable region in phase space, we believe this region deserves a more mathematical description. 
%There should be a way of mathematically characterizing the approximate aperture that we have been investigating in this paper.

The perturbation theories we have been using are asymptotic series and do badly after a certain order. One way to incorporate higher order terms is by resumming the series these expansions generate. It will be useful and interesting to explore methods to resum the series that the BCH and NF methods generate in a way which is able to capture both the presence of resonances and also the presence of the aperture.

We aim to extend our escape rate calculations to higher dimensions. Arnol'd diffusion is another aspect of high-dimensional chaotic motion which we have not addressed here. Several recent analytical methods exist to try and estimate the time scale of Arnol'd diffusion which utilize the multiple invariants mentioned previously \cite{giorgilli}. Alternatively, this time scale can also be estimated by more direct methods \cite{georg}. It would be interesting to examine the competition between the two time scales of ordinary and Arnol'd diffusion in different parts of phase space giving a much more comprehensive picture of the escape process. 

\section*{Acknowledgments}

AR, DLR, AW and JPS were supported by the U.S. National Science
Foundation under Award PHY-1549132, the Center for
Bright Beams. SC was supported by the ARO-MURI Non-equilibrium Many-body Dynamics Grant No. W9111NF-14-1-0003. We thank David Sagan and Alex Dragt for helpful comments. 
%People have been generalizing the method we propose to try and find invariants in higher dimensions. There are several choices available when trying to find different invariants in higher dimensions. A group of people have been fitting polynomials of high orders to trajectories generated by maps where perturbation theory fails. Another group of people have been fitting generating functions and hence generating action angle variables rather than just invariants. The advantage of generating action angle variables is that it allows us to talk of three different energies. In general, these three different energies have different temperatures and hence it is interesting to try and think of the flow of temperature between different energies. 

\bibliographystyle {unsrt}
\bibliography{kickedmap4}

\pagebreak
\widetext
\begin{center}
\textbf{\large Supplemental Materials: Finding stability domains and escape rates in kicked Hamiltonians}
\end{center}

\setcounter{equation}{0}
\setcounter{figure}{0}
\setcounter{table}{0}
\setcounter{page}{1}
\setcounter{section}{0}
\makeatletter
\renewcommand{\theequation}{S\arabic{equation}}
\renewcommand{\thefigure}{S\arabic{figure}}
\renewcommand{\bibnumfmt}[1]{[S#1]}
\renewcommand{\citenumfont}[1]{S#1}

\section{Details of figures}

Here we give a few more details on some of the figures in the paper. The saddle points for the 1-d map given in the main text are calculated by truncating the perturbation theory at a certain order. We have truncated the BCH Hamiltonian to second order. The next order correction does a worse job of capturing the dynamic aperture as can be seen from Figure~\ref{ham2}. We have truncated the NF expansion at 3rd order. If we keep the next order term, the Hamiltonian given below no longer has a saddle point in the region where we expect the boundary of the aperture to be as is evident from Figure~\ref{ham2}. The next order Hamiltonians in the two cases are given by:

\begin{figure}[ht]
\centering
  \begin{subfigure}[b]{.49\linewidth}
    \centering
    \includegraphics[width=.99\textwidth]{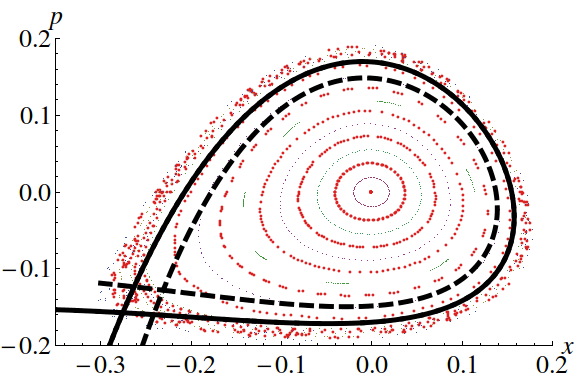}
    \caption{}\label{supfigNF}
  \end{subfigure}%   
  \begin{subfigure}[b]{.49\linewidth}
    \centering
    \includegraphics[width=.99\textwidth]{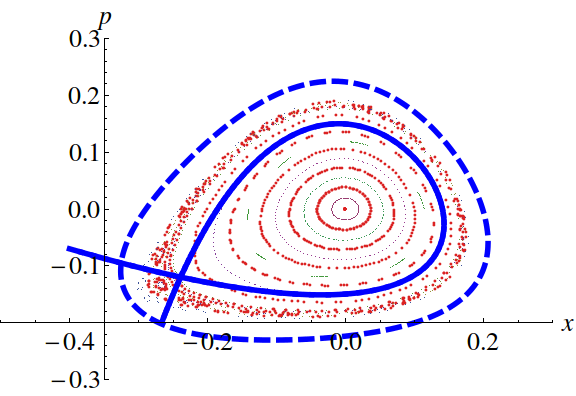}
    \caption{}\label{supfigBCH}
  \end{subfigure}

	\caption{Here we show the effect of including the next order term in both the BCH and NF series. (a) A comparison of the 2nd (black line) and 3rd order (black dashed line) Hamiltonians obtained using BCH. The aperture obtained using the 3rd order Hamiltonian is smaller in size (b) the 3rd order (blue line) NF Hamiltonian has an aperture but the 4th order (blue dashed line) Hamiltonian no longer has a saddle point close to the boundary of the actual aperture as the given contour shows. }
\label{ham2}
\end{figure}

\begin{figure}[ht]
\begin {center}

		\includegraphics[width = 0.45 \textwidth]{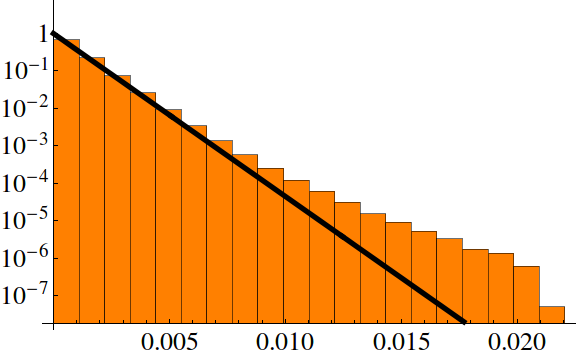}
	\caption{A histogram of the harmonic energy on a logarithmic scale of the particles along with the straight line corresponding to the Boltzmann distribution. The harmonic energy does not describe the ends of the distribution.}
\label{energyhistogramharm}
\end {center}
\end{figure}

\begin{align} 
H_{\textrm{NF}} &= \frac{1}{2} (p^2+x^2 \omega ^2)+\frac{1}{2} \left(-\frac{K^2 p^4 \cos (\omega \tau) (\omega +\tan (\omega \tau))^2}{8 \left(\omega ^3+\omega \right)^2 (2 \cos (\omega \tau)+1)}-\frac{K^2 p^2 x^2 \sec (\omega \tau) \left(\omega^4+\left(3 \omega ^4+1\right) \cos (2 \omega \tau)+4 \omega ^3 \sin(2 \omega \tau)-1\right)}{16 \left(\omega ^3+\omega \right)^2 (2 \cos (\omega \tau)+1)}\right) \nonumber \\
&-\frac{1}{2} \left(\frac{K^2 x^4 \sec (\omega \tau) \left(2 \omega ^3 \sin (2 \omega \tau)- \omega^2-\left(3 \omega^2+1\right) \cos (2 \omega \tau)-1\right)}{16 \left(\omega^2+1\right)^2 (2 \cos (\omega \tau )+1)}+\frac{K p^2 x \sin(\omega \tau)}{2 \omega +4 \omega  \cos (\omega \tau )}+\frac{1}{2} K p x^2\right) , \nonumber \\ &+\frac{1}{2} \left(\frac{K x^3 \omega  \left(\left(\cos^3(\omega \tau)+1\right) \cot (\omega \tau)+\sin (\omega \tau) \cos ^2(\omega \tau)\right)}{4 \cos (\omega \tau)+2}+\frac{K^2 p x^3 \sin (\omega \tau)}{4 \omega +8\omega  \cos (\omega \tau)}\right) + \mathcal{O}(K^3)
\end{align}
\begin{align}
 H_{\textrm{BCH}} &= \frac{1}{2} \left(x^2 \omega ^2+p^2 \right) + \frac{K^2 \tau  x^4-2 K x \left(x^2 \left(\tau ^2 \omega ^2-4\right)-6 p \tau  x-2 p^2 \tau ^2\right) - 2 K^2 \tau^2 p x^3}{48 \tau }
\end{align}

\noindent The 2-d map is a generalization of the sextupole map to higher dimensions. It is given by: 
\begin{align}
 x_{n+1}  &= x_n \cos \omega_1 \tau + \frac{px_n}{\omega_1} \sin \omega_1 \tau , \\
 px_{n+1} &= px_n \cos \omega_1 \tau - \omega_1 x_n \sin \omega_1 \tau - \frac{K}{2} (x_{n+1}^2-y_{n+1}^2) , \\
 y_{n+1}  &= y_n \cos \omega_2 \tau + \frac{py_n}{\omega_2} \sin \omega_2 \tau , \\
 py_{n+1} &= py_n \cos \omega_2 \tau - \omega_2 y_n \sin \omega_2 \tau + K y_{n+1} x_{n+1} 
\end{align}
%\pagebreak
The effective Hamiltonian from BCH can be calculated in the same way as given in the main text. 

The contours are plotted by setting $I_x + I_y = c$ where $c$ is the saddle-point value of $I_x + I_y$. To find the curve that sets the boundary in the space of two invariants, consider setting one of the invariants to a constant and then finding the value of the other invariant which goes through a saddle point. This can be written using a Lagrange multiplyer
\begin{align}  
 \frac{d I_x}{d x} - \lambda \frac{d I_y}{d x} &= 0, \\
 \frac{d I_x}{d px} - \lambda \frac{d I_y}{d px} &= 0 \\
 \frac{d I_x}{d y} - \lambda \frac{d I_y}{d y} &= 0, \\
 \frac{d I_x}{d py} - \lambda \frac{d I_y}{d py} &= 0 \\
 I_y &= c,\\
\end{align}
This gives us $5$ equations with $5$ unknowns $x, px, y, py, \lambda$. These equations were solved numerically and the set of solutions that are closest to the observed numerical boundary are plotted in the main text.

The emittance histogram is drawn by sampling the effective energy by starting at the centre and running for a long time with the kicked noisy map. The emittance is often estimated using a harmonic approximation to the energy. We show in Figure~\ref{energyhistogramharm} that this approximation does well near the centre but does not accurately describe the ends of the distribution.

\end{document}